\documentclass[twocolumn,prb,showpacs,10pt,superscriptaddress]{revtex4-1}
\usepackage{textcomp}
\usepackage{amsmath}
\usepackage{amssymb}
\usepackage{subfig}

\usepackage{color}
\usepackage{multirow}
\usepackage{float} 
\usepackage{fontenc}
\usepackage{graphicx}
\usepackage{graphics}
\usepackage[justification=centering]{caption}
\usepackage{lipsum}
\makeatletter
\newcommand*{\rom}[1]{\expandafter\@slowromancap\romannumeral #1@}
\newcommand{\Rmnum}[1]{\expandafter\@slowromancap\romannumeral #1@}
\usepackage{hyperref}

\begin{document}
\title{A Density Functional Theory based study of Electron Transport Through Ferrocene Compounds with Different Anchor Groups in Different Adsorption Configurations of A STM-setup}
\author{Xin Zhao}
\affiliation{Institute for Theoretical Physics, TU Wien - Vienna University of Technology, Wiedner Hauptstrasse 8-10, A-1040 Vienna, Austria}
\author{Robert Stadler}
\email[Email:]{robert.stadler@tuwien.ac.at}
\affiliation{Institute for Theoretical Physics, TU Wien - Vienna University of Technology, Wiedner Hauptstrasse 8-10, A-1040 Vienna, Austria}

\begin{abstract}
In our theoretical study where we combine a nonequilibrium Green's function (NEGF) approach with density functional theory (DFT) we investigate compounds containing a ferrocene moiety which is connected to i) thiol anchor groups on both sides in a para-connection, ii) a pyridyl anchor group on one side in a meta-connection and a thiol group on the other side in a para-connection, in both cases with acetylenic spacers in between the Ferrocene and the anchors.
We predict possible single molecule junction geometries within a scanning tunneling microscopy (STM) setup
where we find that the conductance trend for the set of conformations are intriguing in the sense that the conductance does not decrease while the junction length increases which we analyze and explain in terms of Fermi level alignment. We also find a pattern for the current-voltage (IV) curves within the linear-response regime for both molecules we study, where the conductance variation with the molecular configurations is surprisingly small.
\end{abstract}

\maketitle
\subsection{Introduction}

The field of molecular electronics has been active for more than 40 years since Aviram and Ratner originally proposed a single molecule as a device, namely a molecular rectifier ~\cite{Aviram}. Since the mid 1990s experimental advances for the electronic characterization of molecular junctions \cite{Agrait, Nitzan, Xu,Eigler} helped to promote the  field of single molecule electronics which aims at maintaining a continuous rise in performance of digital devices even once the lower threshold for miniaturization faced by the semiconductor industry has been reached. Molecules can self-organize when adsorbed to electrodes, which is an essential advantage for overcoming technological difficulties. For making single-molecule junctions applicable as molecular devices, they need to be operable at room temperature. Experimentally, these ambient conditions can
be achieved with an electrochemical scanning tunneling microscopy (E-STM) \cite{Tim2006, Tim2005, Haiss, Ricci}.
In such experiments, the target molecules are adsorbed on the metal surface of the working electrode. When a STM tip is brought to the solution containing target molecules and withdrawn after that, in some cases a molecule is trapped by the tip and the contacted metal surface and thereby a junction is formed. During the approaching and withdrawing process different geometrical contacts can be accessed, which increases the difficulty for the quantitative description in theoretical simulations.  
For a theoretical description of electron-transport in a single molecule junction usually
a non-equilibrium Green's function formalism (NEGF) ~\cite{keldysh} combined with density functional theory (DFT) is used.
The simulation of molecular electron transport properties enables a clearer understanding of the conductance dependence on the molecular structures and helps the design of logical gates ~\cite{Stadler2004} and transistors ~\cite{Tim2005} in single molecule electronics as well as the implementation of thermoelectric devices ~\cite{fano,lambert2}.
\begin{figure}
    \captionsetup[subfigure]{labelformat=empty}
    \centering
    \subfloat[]{%
     \includegraphics[width=\linewidth]{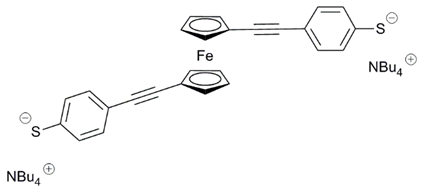}%
     }

    \subfloat[]{%
     \includegraphics[width=\linewidth]{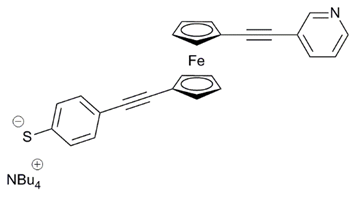}%
     }
     \caption{\small a) s-FDT with a phenylthiolate group on each side; b) s-FPT with a pyridyl group on one side and thiolate on the other side.}
 \label{Fig.chemicalstru-sFXT}
\end{figure}

We investigate the electron transport properties of two type of molecular wires containing a ferrocene moiety with the same and different anchor groups ~\cite{Tim2013}, which are shown in Fig. ~\ref{Fig.chemicalstru-sFXT}. Here s-FDT is a short notation for 1,1'-bis(thiophenol-4-ethynyl)ferrocene with phenylthiolate groups on both ends of the molecule and s-FPT for 1-(3-pyridylethynyl)-1'-(thiophenol-4-ethynyl)ferrocene where one of the anchor groups is substituted by pyridyl.
The special characteristics of these two molecules are: i) the ferrocene moiety can be oxidized and therefore enables switching between two redox states, which can be operated via gating ~\cite{Tim2006}, ii) the molecular backbone is the same for both cases while the different conductance behavior for the two molecules due to the difference in anchor groups is a focus in our study. Both pyridyl groups ~\cite{wong,pyridil1,pyridil2,pyridil3,pyridil4} and thiolate ~\cite{Kim, Beebe, Liang} are popular choices for anchor groupss in the single molecule electronics community. In our molecular design the acetylenic spacers are added to increase the distance between the two electrodes in the molecular junction for the prevention of through-vacuum tunneling and for the separation of the redox-active centers from the leads. 

The paper is organized as follows: In Sec. ~\ref{paper4:junction-geos} we present possible configurations for the junction. In Sec. ~\ref{sec:Computational_details} we outline the computational details of our approach. In Sec. ~\ref{paper4:lenth-dependence} we identify a pattern for the conductance in dependence on the junction geometry where for each molecule there is a range of higher conductance values and a single lower value. In our calculations the lower value is obtained form a linear junction geometry and the higher ones correspond to bended configurations. The conductance for the bended configurations increases with an increase in length for s-FDT and oscillates for s-FPT, which is counter intuitive in the coherent electron transport regime. 
In Sec.~\ref{paper4:analysis} we analyze this trend in terms of Fermi level alignment and we conclude with a brief summary of our results in Sec.~\ref{paper4:summary}.

\subsection{Choice of geometries}\label{paper4:junction-geos}

Measurements in a E-STM setup are operated under ambient conditions, where thiolates are likely to be saturated with hydrogen atoms even if they are adsorbed on a metal surface as salts.
Since it has also been shown that the dissociation of hydrogen from a thiol group thereby forming a thiolate is not energetically favorable during the adsorption of this anchor group on a gold surface ~\cite{Silva, Sanvito1, Nara, Guo}, we keep the hydrogen bonded to the sulfur in all calculations presented om this work.

\begin{figure}
\begin{center}
\includegraphics[width=1.0\linewidth,angle=0]{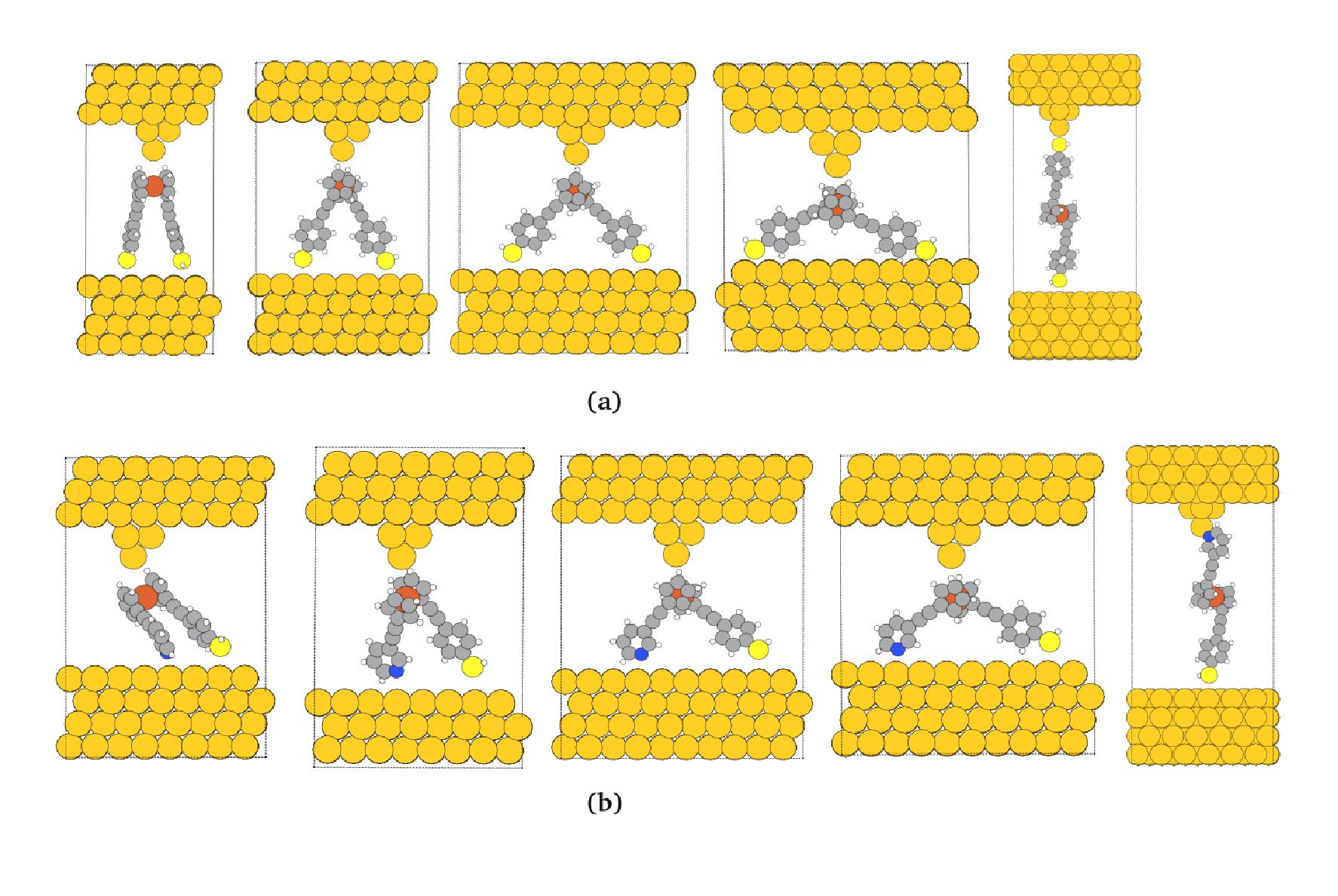}
\caption{\small Junction geometries for a) s-FDT  and b) s-FPT with different rotation angles.\label{Fig:geo}}
\end{center}
\end{figure}

\begin{figure*}
\begin{center}
\includegraphics[width=1.0\linewidth,angle=0]{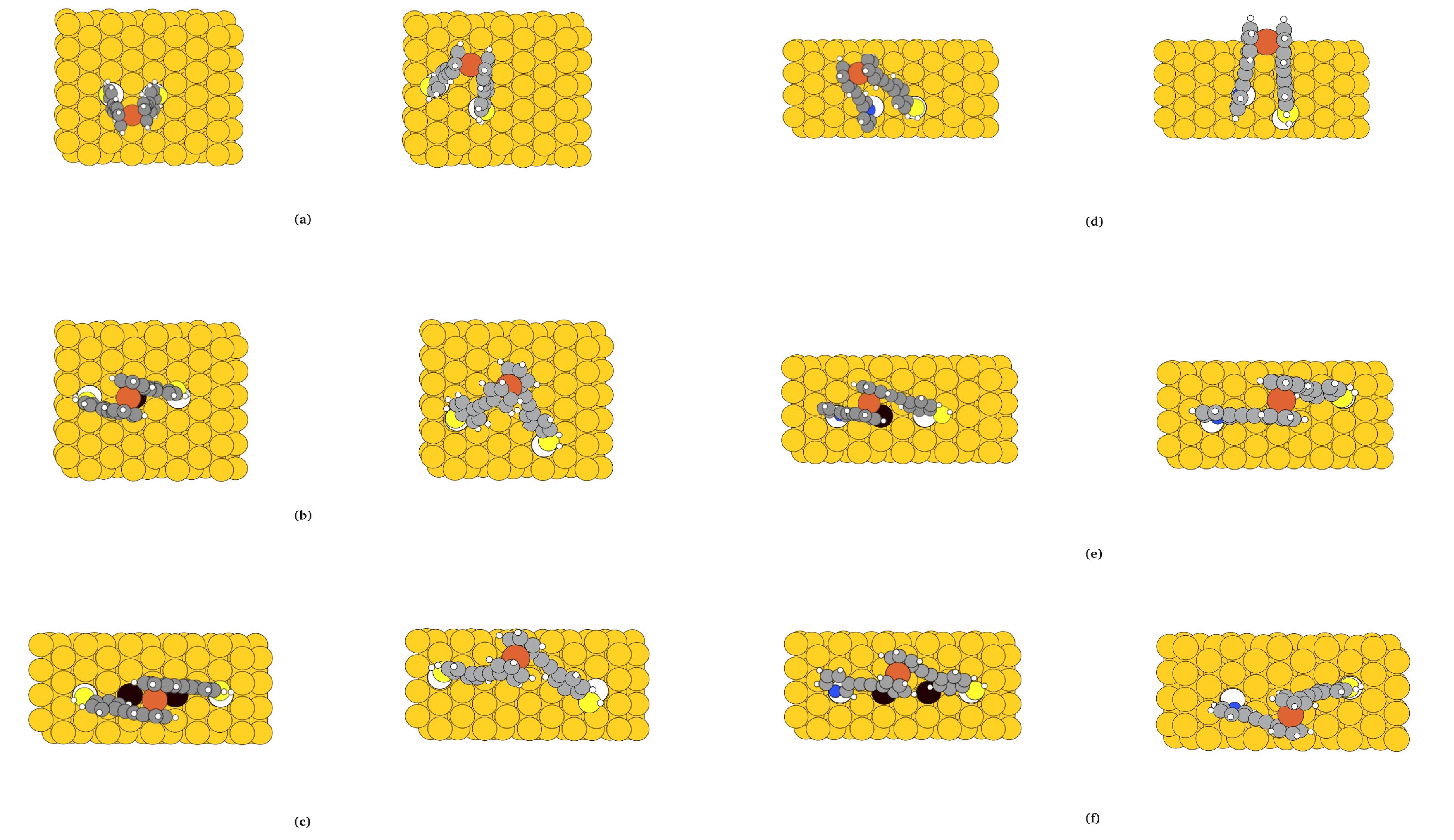}
\caption{\small Adsorption geometries for same sublattice (left panels) and different sublattice adsorption (right panels) for the molecules s-FDT (a-c) and s-FPT (d-f), where the distances d(Au-S) between the respective sulfur and gold sites resulting from on-top adsorption of both anchor groups are a) 3.32 \AA{}, b) 3.17 \AA{}, c) 3.18 \AA{} for s-FDT and in all cases are identical for same (left panels) and different (right panels) sublattice adsorption, while for the asymmetric s-FPT d(Au-S)/d(Au-N) were found to be d) 3.30/2.76 \AA{} (left panel),  3.11/3.14 \AA{} (right panel), e) 3.60/3.01 \AA{} (left panel), 3.70/3.13 \AA{} (right panel), f) 3.55/2.74 \AA{} (left panel), 3.55/2.85  \AA{} (right panel) by our optimization procedure as described in the main text. The contacted gold positions on the surface for each adsorption geometry are marked in white, while the Au atoms on the same fcc sublattice in between these positions are marked in black for geometries with both anchors adsorbed on the same sublattice.}\label{Fig.bendedsublattice}
\end{center}
\end{figure*}

During the approaching and withdrawing process of the tip in E-STM experiments, different geometries for the established molecular contact can occur even if there is only one type of molecule in the solution. In principle, one could imagine three possible sources for the variation of the conductance in dependence of the junction structure for any particular molecule: i) inner (e.g. conformational) degrees of freedom of the molecular structure, ii) variations in the surface or tip structure occurring during the repeated measurement process or iii) differences in the adsorption geometry of the molecules in the junction. While variations in the surface structure of the working electrode itself are likely to occur while conductance data is recorded, there is no indication that such gradual variations would result in distinct peaks in histograms. The inner degrees of freedom of the molecule itself, most notably the rotational angle related to the two anchor groups attached to the Ferrocene (Fc) moiety, on the other hand cannot be identified with distinct local minima in the corresponding total energies where the two cyclopenta-dienyl (Cp) rings are known to rotate quite freely with energy barriers of only a few kJ/mol. So we assume that the defining element for the likelihood of particular junction geometries to be formed lies in the commensurability of the rotation angle between the anchor groups of the molecule with the sequence of potential on-top gold positions defined by the fcc (111) surface structure. This selection criterion leaves only few potential adsorption geometries for both s-FDT and s-FPT, where one is linear in the sense that one of the anchor groups is attached to the surface and the other one to the tip, while for the others both anchor groups bond to the surface but the rotation angle varies within the lower limit of being zero and the upper limit of the Fc moiety touching the surface, where the Fc is contacted directly by the tip in this setup. A range of potential junction geometries are shown in Fig. ~\ref{Fig:geo}.
In this section we provide further details on the step-wise structure optimization process we adopted for the definition of all the relevant degrees of freedom including the distances between the respective molecules and the tip and surface.
For the linear structures with one anchor group attached to the substrate and the other one to the tip, we assumed a rotation angle of 180 \textdegree and the sulfur or nitrogen atom of the contact site on the molecule to be adsorbed on top of a respective gold position. For the structures with both anchor groups adsorbed on the surface and the tip directly contacting the Fc- moiety, the rotation angle was defined by the criterion that the sulfur or nitrogen contact atoms of both anchor groups needed to be on-top of two gold atoms of the surface, without the Fc moiety touching the surface. This criterion gives two types of potential Au contact positions, i.e. the anchoring atoms are either adsorbed to the same fcc sublattice or different sublattices as shown in Fig. ~\ref{Fig.bendedsublattice}, where the number of Au atoms in between the two contacted sites are 
zero (Fig. ~\ref{Fig.bendedsublattice} a),d)), one (Fig. ~\ref{Fig.bendedsublattice} b),e)) or two (Fig. ~\ref{Fig.bendedsublattice} c),f)) on a line.
\begin{figure}
\begin{center}
\includegraphics[width=1.0\linewidth,angle=0]{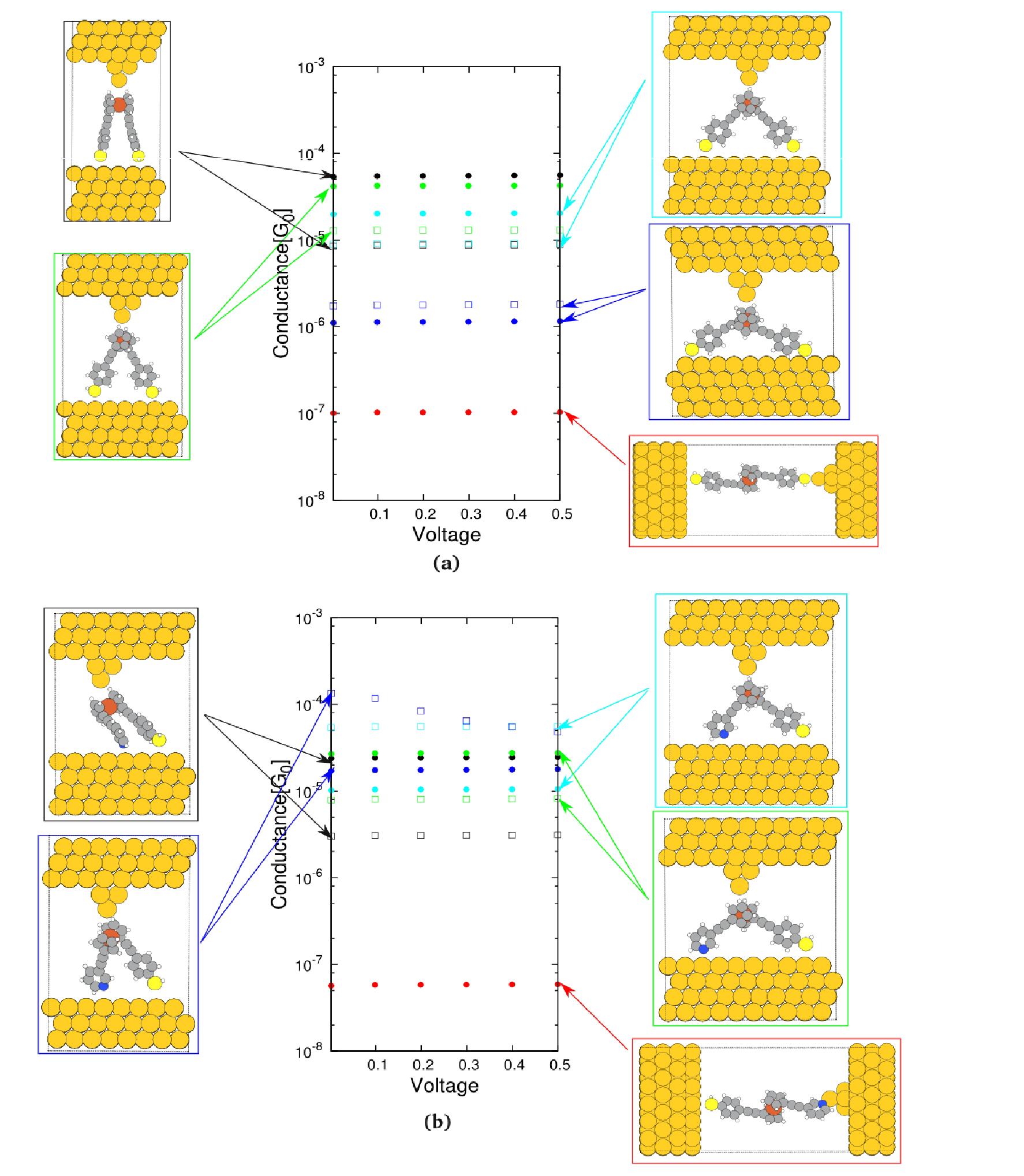}
\caption{\small Conductance in dependence on the voltage from DFT calculations with SO corrections, where dots in each panel represent the anchoring sites of molecules adsorbed on the same sublattice on the surface, and squares represent anchoring sites on different sublattices for selected junction geometries of a) s-FDT with a rotation angle of 0\textdegree (black),  54\textdegree (green), 89 \textdegree (cyan), 125 \textdegree (blue) and 180\textdegree (red); b) s-FPT with a rotation angle of 0\textdegree (black), 50\textdegree (blue), 92\textdegree (cyan), 117\textdegree (green) and 180\textdegree (red).\label{Fig:IV}}
\end{center}
\end{figure}
For obtaining stable junction geometries from the NEGF-DFT calculations, we i) relaxed all nuclear positions on the molecules in a linear conformation by means of total energy minimization, ii) adjusted the rotation angle between anchor groups manually so that it fit the commensurability requirement with the surface illustrated in Fig. ~\ref{Fig.bendedsublattice}, iii) relaxed all nuclear positions on the molecule again, and then iv) optimized the distances between the respective contact atoms on the electrodes and on the molecules. This distance optimization resulted in d$_{Au-Fe}$ = 4.45 \AA{} for the contact of the Fc moiety to the lowest atom on the tip for all adsorption geometries depicted in Fig. ~\ref{Fig:geo}, while d$_{Au-S}$ and d$_{Au-N}$ for the respective contacts of the anchor groups to the substrate varied more widely for each configuration and the values we obtained are explicitly given in the caption of Fig. ~\ref{Fig.bendedsublattice}. The respective linear adsorption geometries for both molecules have one anchor group contacting the substrate with d$_{Au-S}$ $=$ 3.08 \AA{} for s-FDT and 3.12 \AA{} for s-FPT, and the other anchor group contacting the tip with d$_{Au-S}$ $=$ 2.61 \AA{} for s-FDT and d$_{Au-N}$ $=$ 2.21 \AA{} for s-FPT, respectively.

\subsection{Theoretical/Computational methods}\label{sec:Computational_details}

We performed DFT- calculations with a PBE XC-functional within a NEGF framework ~\cite{Brandbyge, Xue, Rocha} using the GPAW code ~\cite{Mortensen, Enkovaara} to compute transmission probabilities $\mathcal{T}(E)$. In order to account for self-interaction errors and image charge effects present in DFT with semi-local XC-functionals we applied a scissor operator (SO) correction according to Quek et al. ~\cite{Quek}. 
This ad hoc correction is only applicable for rather weakly coupled molecules, a condition which is fulfilled for dithiols but not dithiolates adsorbed on gold surfaces ~\cite{deMeloSouza}. In Ref.~\cite{deMeloSouza} it was demonstrated that the dissociative adsorption of dithiols on Au substrates is energetically highly unfavourable and that for adsorbed dithiols a good agreement with experimental data for the conductance can be achieved when the values calculated from NEGF-DFT are corrected with a SO approach ~\cite{Thygesen}, which we briefly summarize also in Appendix ~\ref{Appe1}. 
All DFT calculations were carried out without treating spin polarization as a degree of freedom because previous studies on similar Ferrocene complexes ~\cite{Xin2017} revealed the low spin configuration to be the ground state. 
The I-V curves we simulated for the geometries of each molecule (as shown in Fig. ~\ref{Fig:geo}) are obtained from the transmission functions $\mathcal{T}(E)$ in a rigid band approximation where the bias dependence of $\mathcal{T}(E)$ is disregarded. This means that the current is determined within the linear response regime as an integration over the zero bias transmission function, and the conductance is simply calculated as
 G(V)= I(V)/V
 \label{G(V)}
for each voltage V.


\subsection{Voltage and structure dependence of the conductance}\label{paper4:lenth-dependence}

In Fig. ~\ref{Fig:IV} we show the voltage dependence of the conductance for all junction structures for both s-FDT and s-FPT, where we can identify the following trends: i) the quantitative values for both molecules are rather similar, and there is a pattern for each case with a range of higher conductance values and a single lower value, where in our calculations the lower value corresponds to the linear junction geometry and the set of higher ones to the bended configurations, i.e adsorption structures with both anchor groups attached to the substrate, the conductance varies only within a range of two orders of magnitude for the bended structures of both molecules; ii) the conductance for all bended configurations increases with an increase in length for s-FDT and oscillates for s-FPT, which is counter intuitive for the coherent electron transport regime where the tunnel current is expected to increase with a decrease of the junction length as we discuss below; iii) the conductance dependence on the voltage (V) is not very significant, i.e. the conductance does not change as much with V as it would be expected even under the rigid approximation we make by neglecting the influence of V on $\mathcal{T}(E)$.
\begin{figure*}
 \begin{center}
\includegraphics[width=1.0\linewidth,angle=0]{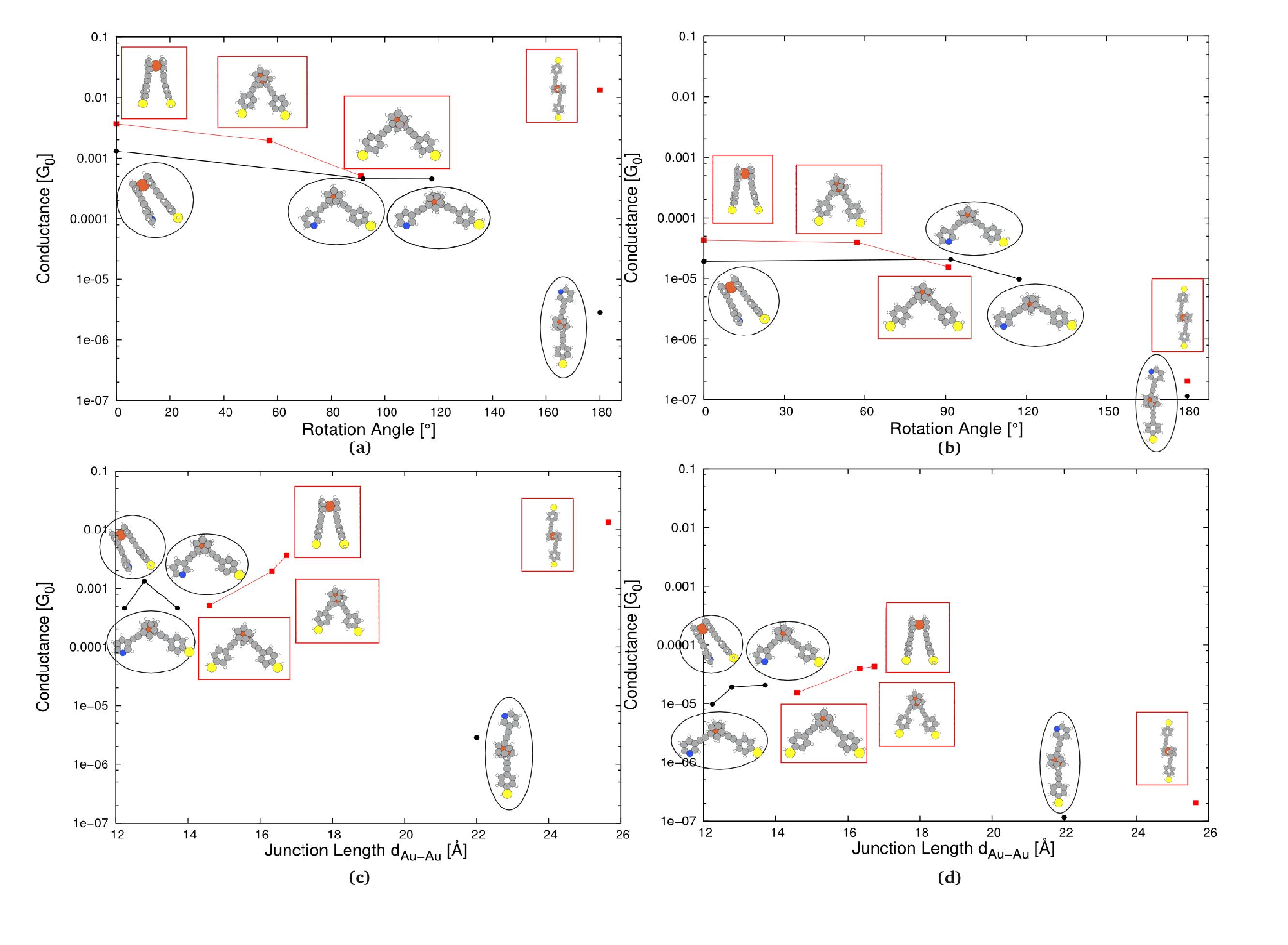}
\caption{\small Zero-bias conductance for a selected range of structures as calculated from $\mathcal{T}(E_{F})$ with NEGF-DFT a),c) without and b),d) with SO corrections in dependence on a),b) the rotation angle between anchor groups and c),d) the junction length defined as the distance between the surface plane of the substrate and the contact Au atom on the tip.\label{Fig.conductance_vs_angle}}
\end{center}
\end{figure*}

\begin{figure*}
\begin{center}
\includegraphics[width=1.0\linewidth,angle=0]{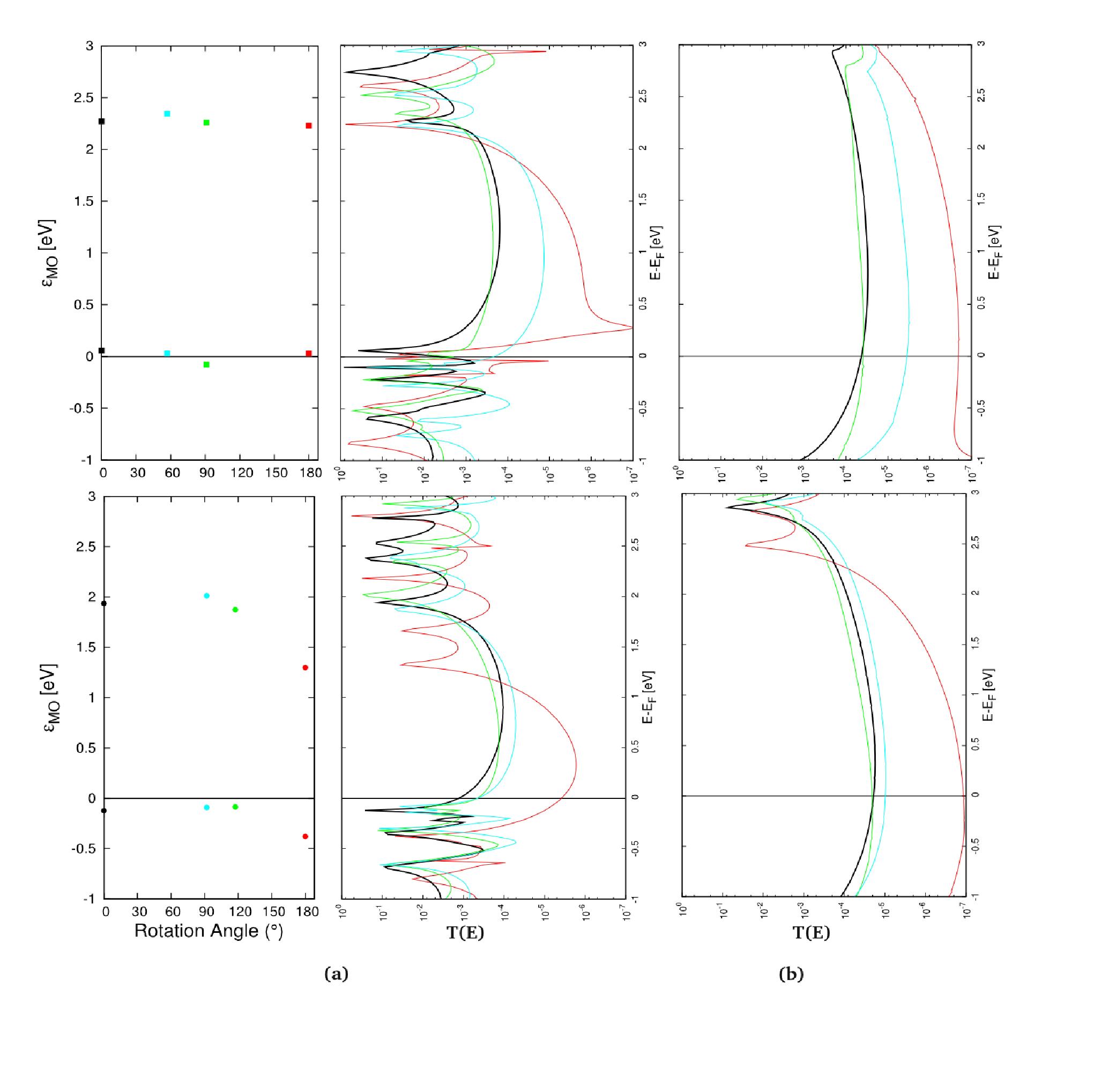}
\caption{\small Eigenenergies of the HOMO and LUMO from a subdiagonalization of the molecular part of  the transport Hamiltonian for the junction setups in Fig. ~\ref{Fig.conductance_vs_angle} (a) left panels), transmission functions $\mathcal{T}(E)$ without ((a) right panels) and with SO corrections (panel b)), where the upper panels are for s-FDT and the lower panels for s-FPT. The colors attributed to the respective four junction geometries are the same as those chosen for Fig. ~\ref{Fig:IV}. \label{Fig:HOMOandtf}.
\label{Fig.tf_noSO}}
\end{center}
\end{figure*}

In Fig. ~\ref{Fig.conductance_vs_angle} we selected the linear conformation and some bended configurations with both anchor groups bonded to Au atoms with the same fcc sublattice and focus on exploring the reason behind the observed conductance trend.
We plot the zero bias conductance we obtain directly in $\mathcal{T}(E_{F})$ from the NEGF-DFT calculations in dependence on the rotation angle between the anchor groups attached to the Fc moiety as well as on the junction length which we define as the distance between the surface plane of the substrate and the position of the tip atom contacting the molecule from above. While usually the conductance decreases with the junction length for electron transport in the coherent tunneling regime, we find for the conductance without SO corrections that it even increases in our calculations for s-FDT and oscillates for s-FPT when both anchor groups are attached to the substrate as illustrated in Fig.~\ref{Fig.conductance_vs_angle}. The SO correction opens up the HOMO-LUMO gap for all geometries (Fig. ~\ref{Fig.tf_noSO}), thereby reducing the conductance values quite substantially but it should not change the fundamental reasons for the conductance trend that we observe in Figs. ~\ref{Fig:IV} and ~\ref{Fig.conductance_vs_angle}. The conductance without SO corrections directly correlates with the energetic positions of the HOMO $\varepsilon_{HOMO}$ (Fig. ~\ref{Fig.tf_noSO}), which is defined by a delicate angle dependent Fermi level alignment, where both the component of the molecular dipole moment in the transport direction and differences in zero bias charge transfer play decisive roles as we will demonstrate in a detailed analysis of the structure dependent interplay of the two effects in section ~\ref{paper4:analysis}.

\subsection{Analysis of the Dependence of the Conductance on the Junction Length in Terms of Fermi Level Alignment}\label{paper4:analysis}
   
\begin{figure*}
\begin{center}
\includegraphics[width=1.0\linewidth,angle=0]{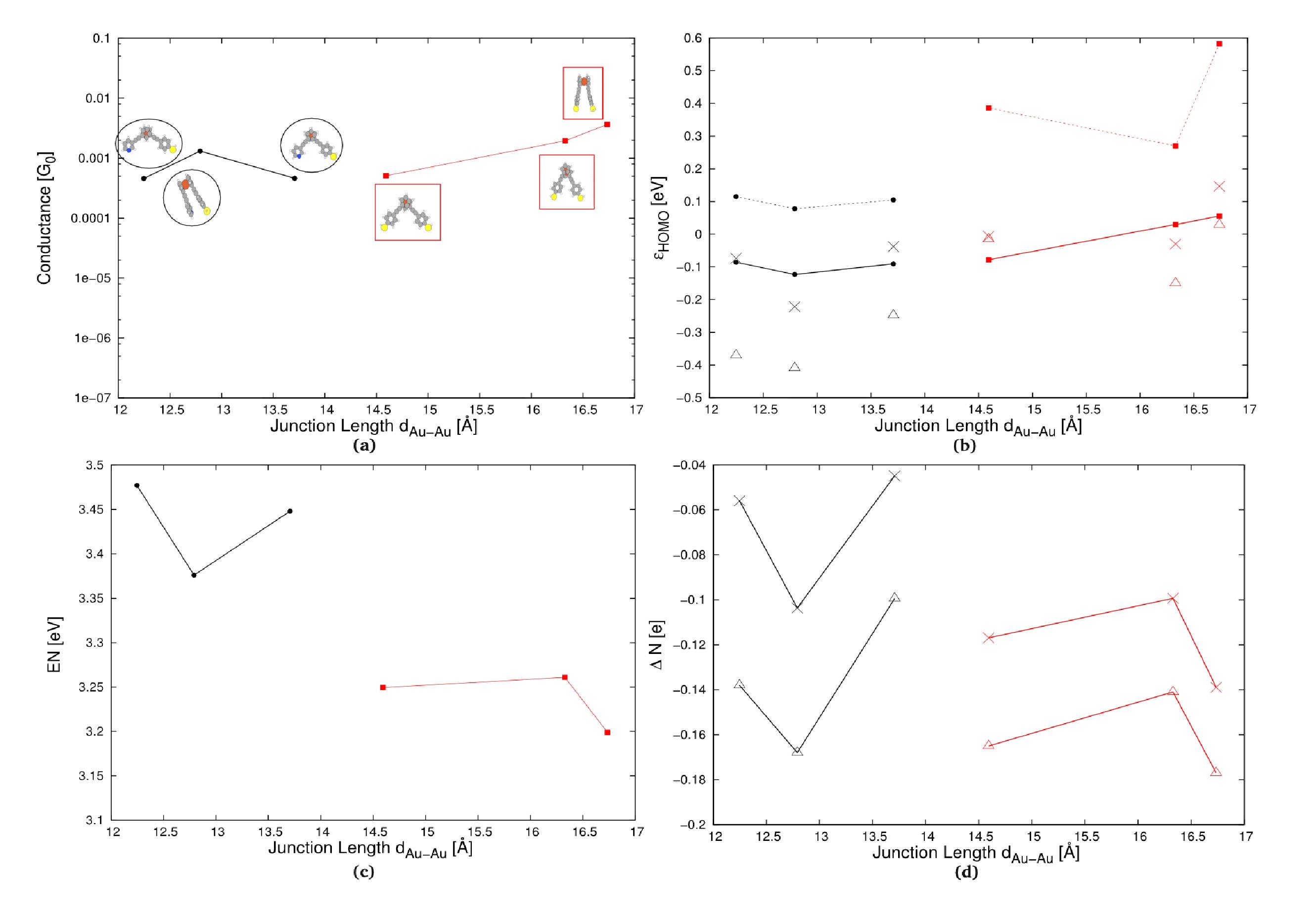}
\caption{\small Junction length dependence of a) the conductance without SO corrections, b) the eigenenergy of the HOMO from a subdiagonalization of the molecular part of the transport Hamiltonian of the junction (solid) and Vacuum level alignment (dotted), where for the latter shifted values taking into account the contribution from partial charging are also shown as crosses (Bader analysis) and triangles (integration over $\Delta$ n(z)), c) Electronegativities of the free molecules in dependence on the rotation angle and d) partial charges on the molecules in the junction from a Bader analysis (crosses) as well as from integration over $\Delta$ n(z) in the molecular region (triangles) in units of electrons. For all panels symbols and lines are shown in black for s-FPT and in red for s-FDT.\label{Fig.conductance_dependence}}
\end{center}
\end{figure*}

The SO correction should not change the trend of the conductance dependence of the junction length, therefore in the following we focus on the NEGF-DFT calculations without SO corrections where in Fig. ~\ref{Fig.conductance_dependence} a) we provide the structure dependent conductance on a smaller molecular length range when compared with Fig. ~\ref{Fig.conductance_vs_angle} c). The energetic positions of the HOMO in Fig. ~\ref{Fig:HOMOandtf} as obtained from a diagonalization of the subspace of the molecular region of the transport Hamiltonian show that the closer $\varepsilon_{HOMO}$ is to the Fermi Level E$_{F}$, the higher the conductance becomes. The only exception from that trend is the slightly tilted cis-configuration of s-FPT where a distinctly broadened peak in the corresponding transmission function (Fig. ~\ref{Fig.tf_noSO} a), black curve) indicates an overall increased electron coupling of the HOMO to the leads in this particular configuration.
This means that any analysis of the conductance trends must focus on the sources of Fermi level alignment (FLA) of MOs obtained from a subdiagonalization in general and the HOMO in particular.
These sources can be divided into properties of the separate components of the junction and the charge rearrangement when these components are combined to form the junction. The former consist mostly of the permanent dipole of the free molecule in the transport direction which has a strong dependence on the contact angle but also the dipole moment of the metallic slab which is caused by the asymmetry that arises from the fact that there is a tip on one side but a flat surface on the other and the variation of the vacuum level positioning with the length of the gap between tip and substrate. These effects result in the dotted lines in Fig. ~\ref{Fig.conductance_dependence} b), which were obtained from vacuum level alignment from independent DFT calculations for the free molecule and the metal slabs without the molecule, where the vacuum level is calculated as the electrostatic potential in the vacuum region for both separate systems and the HOMO of the molecule and E$_{F}$ of the metal slab  put in reference to this vacuum level are assumed to be the same for both when the junction is formed. For the details of this procedure we refer to Appendix ~\ref{Appe2}, where the values for $\varepsilon_{HOMO}$ obtained in this way are all substantially higher in energy than those obtained from the subdiagonalization of the transport Hamiltonian of the junction (solid line in Fig. ~\ref{Fig.conductance_dependence} b)) but show very similar qualitative trends for s-FPT (black lines) and somewhat different behavior for s-FDT (red lines). 
These differences are due to the neglect so far of the effect of charge rearrangements when the components of the junction are brought together. There are different views in the literature of how best to describe this effect, where some authors prefer a description in terms of interface or bonding dipoles ~\cite{Verwuste} and one of us has written a series of papers where they are discussed in terms of partial charges on the molecule ~\cite{Stadler2006, Stadler2007, Stadler2010, Kastlunger2013}. Both views agree in attributing the charge rearrangement at the interface to Pauli repulsion, but while the interface dipole approach does not allow for a quantitative analysis of the resulting energy shifts of MOs ~\cite{Verwuste}, these can be reproduced to a high level of accuracy from vacuum level aligned eigenenergies from free molecule calculations where the partial charge the molecule also has in the junction because of its interaction with the leads is added ~\cite{Stadler2006, Stadler2007, Stadler2010, Kastlunger2013}. 
This is because in case of a positive charge the remaining electrons are bound more tightly to the respective nuclei in the molecule leading to more negative MO eigenenergies, in case of a negative charge the opposite behavior with a rise in MO energies is encountered ~\cite{Stadler2006}. For such a description it is not relevant whether the partial charge comes from the emptying or filling of a molecular level or from Pauli repulsion effects at the interface, where a clear distinction of both in any case is not rigorously possible in the junction setup where the MOs and the surface states of the leads form hybrid orbitals. The only source of accuracy in this partial charge perspective is the definition of this charge in its technical practicality, where the options are a Mulliken analysis ~\cite{Mulliken}, a Bader analysis ~\cite{bader} or an integration over $\Delta$n(z), i.e. the difference of the electron density between the junction and its two components in the transport direction and averaged over the surface plane in the molecular region \cite{Stadler2006, Stadler2007}. 
Since we are interested in the electrostatic effect of the charge on the MOs, a Mulliken analysis where electrons are attributed to atoms in terms of the LCAO basis sets is not a good choice. The Bader analysis which uses zero flux surfaces to divide atoms or the integration over $\Delta$n(z) where the borders on $z$ for the definition of the molecular region are put at the mid-point between the $z$-coordinates of the molecular and metallic contact atoms provide much more useful molecular partial charge values $\Delta$N for this purpose.
In Fig. ~\ref{Fig.conductance_dependence} c) we plot the electronegativities EN of the free molecules according to Mulliken's definition of EN = (IP+EA)/2 ~\cite{Mulliken} where the ionization potential IP and the electron affinities EA have been calculated from DFT total energies of charged and neutral molecules as IP=E(N-1)-E(N) and EA = E(N)-E(N+1), respectively, where N is the number of electrons in the neutral molecules. The junction length dependence of EN in Fig. ~\ref{Fig.conductance_dependence} c) corresponds qualitatively with that of the molecular partial charges $\Delta$N in Fig. ~\ref{Fig.conductance_dependence} d), which were obtained from a Bader Analysis (crosses) and by integrating over $\Delta$n(z) (triangles), where stronger negative values mean that larger fractions of an electron have been subtracted from the molecule due to the formation of the junction. 
These values of $\Delta$N, where the results from the Bader analysis and from the integration disagree quantitatively in the sense that the Bader charges are smaller, have also been used for the vacuum level alignment with partially charged molecules in Fig. ~\ref{Fig.conductance_dependence} b), where the Bader shifted energies seem to show better agreement with $\varepsilon_{HOMO}$ obtained from a subdiagonalization of the transport Hamiltonian of the junctions. For s-FPT both EN and $\Delta$N follow the same trend as the vacuum level aligned $\varepsilon_{HOMO}$ without the consideration of partial charges, while for s-FDT the second point whose value is the lowest one in the vacuum level alignment of the neutral molecules has the highest value in EN and $\Delta$ N, thereby flattening the qualitative trends in the length dependence of the final $\varepsilon_{HOMO}$ Fig. ~\ref{Fig.conductance_dependence} b) and the conductance in Fig. ~\ref{Fig.conductance_dependence} a).

Because the values for $\varepsilon_{HOMO}$ in Fig. ~\ref{Fig.conductance_dependence} b) are quite close to E$_{F}$, the question arises whether Fermi level pinning plays a role, i.e. whether the energetic distance of $\varepsilon_{HOMO}$ to E$_{F}$ can be tuned by variations in EN or will be pinned at the electrodes Fermi level regardless of EN. In Ref. ~\cite{VanDyck} a test for the occurrence or absence of Fermi level pinning was proposed by conducting cluster calculations, where the respective molecule is just attached to single gold atoms at each metal contact. For the molecules in our study this would mean that for the linear structures where one anchor group is bonded to the surface and one to the tip there would be two gold atoms in the cluster calculations, namely one at each anchor group, while the junction geometries with both anchor groups adsorbed on the surface would in addition have a third gold atom, where the Fc moiety is contacted by the tip.
In Ref. ~\cite{VanDyck} it was proposed that when the HOMO is a metal-induced gap state (MIG) in these cluster calculations, i.e. it is mostly localized on the gold atoms, any Fermi level pinning effect is annihilated and the energetic position of the HOMO level is determined by the respective electronegativity of the molecule. 
\begin{figure}
\begin{center}
\includegraphics[width=1.0\linewidth,angle=0]{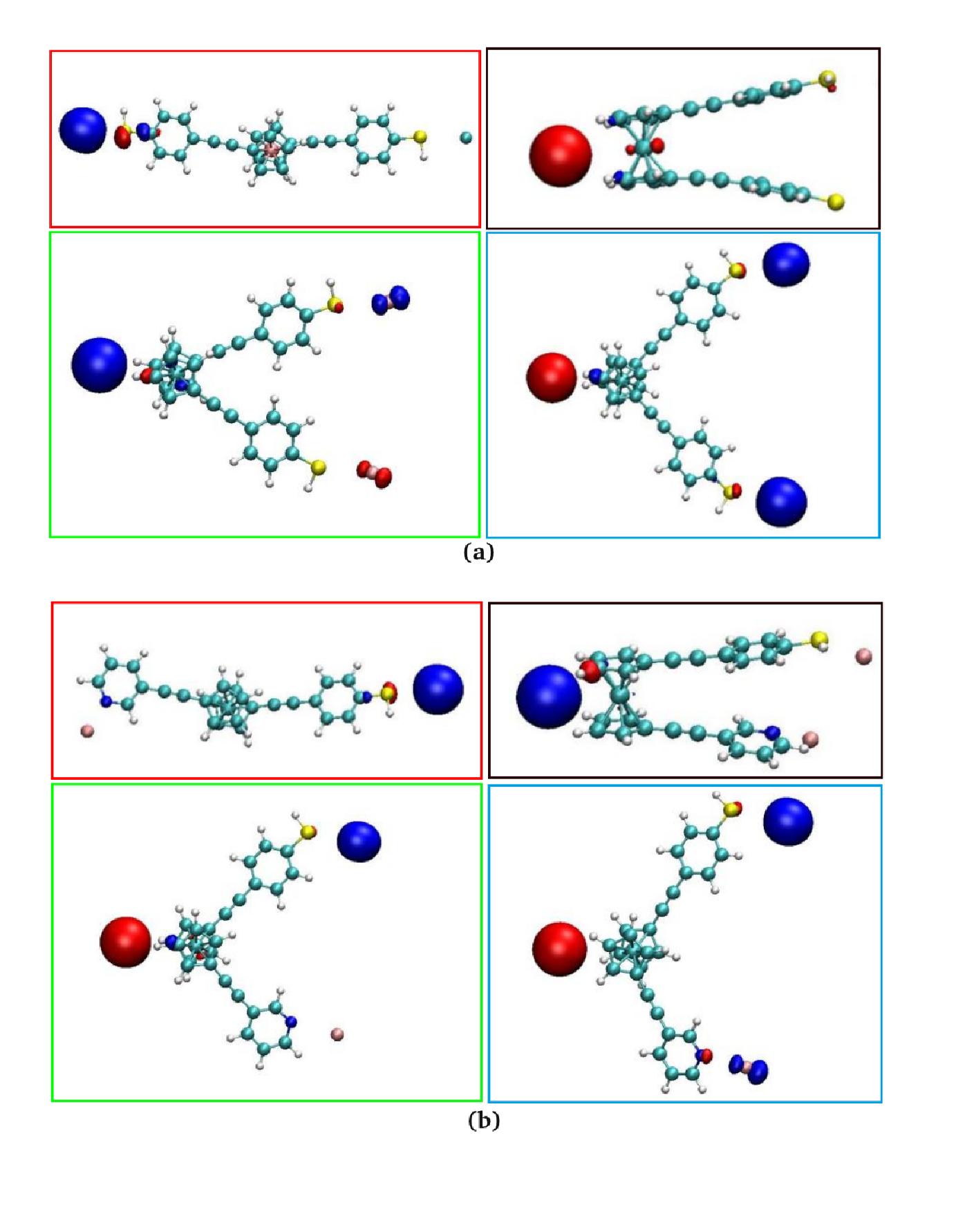}
\caption{\small Localization patterns for the HOMO from cluster calculations containing the free molecule and the two or three Au contact atoms defined by the respective junction geometries for a) s-FDT  and b) s-FPT.\label{Fig.MIGSorb}}
\end{center}
\end{figure}
In Fig.~\ref{Fig.MIGSorb} we plot the respective HOMOs from cluster calculations for the junction geometries discussed in Fig. ~\ref{Fig.conductance_vs_angle}, ~\ref{Fig:HOMOandtf} and ~\ref{Fig.conductance_dependence}. Since for all cases the HOMO has MIG character, Fermi level pinning can be disregarded as a factor in the observed FLA trends.

\begin{figure*}
\begin{center}
\includegraphics[width=1.0\linewidth,angle=0]{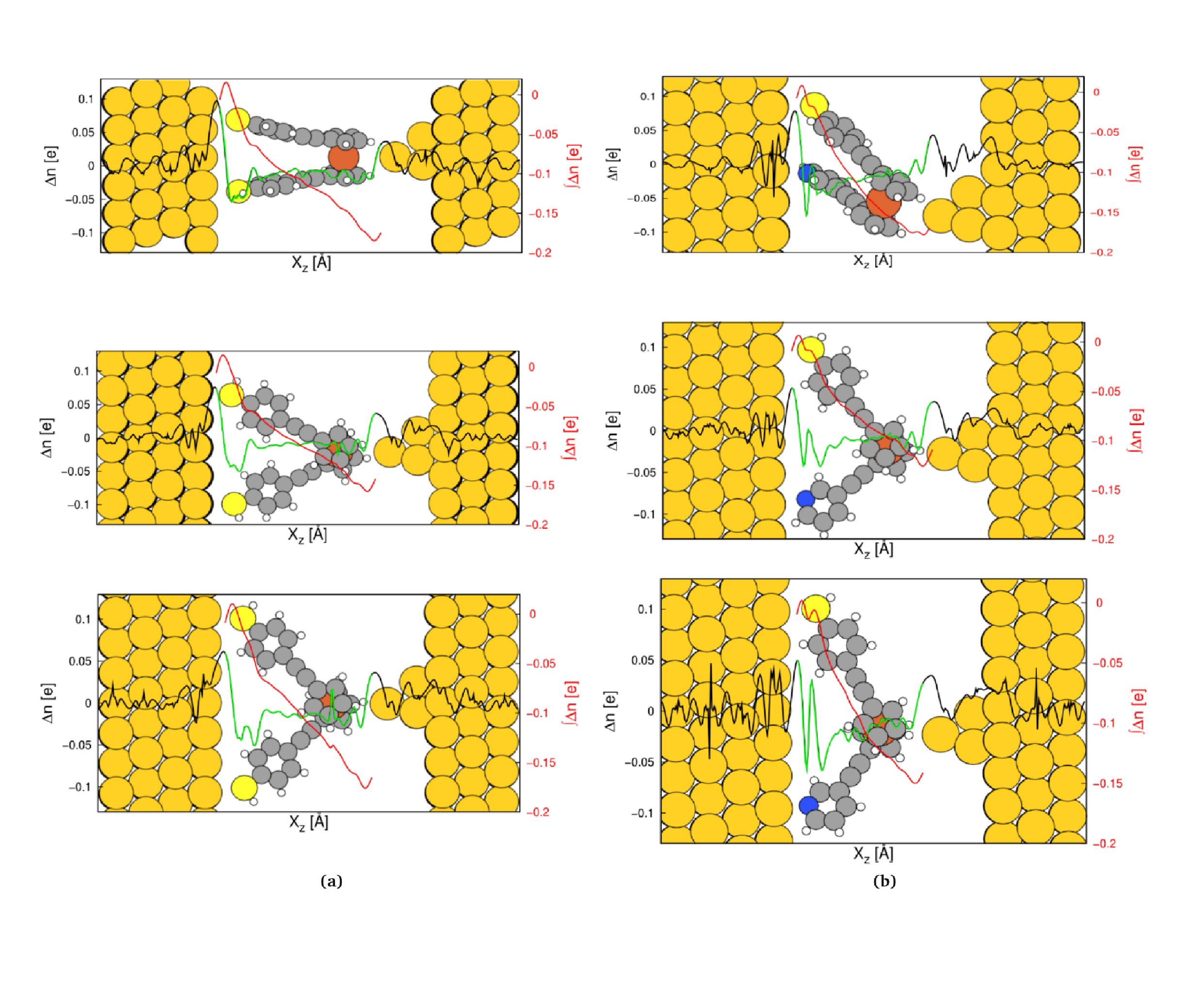}
\caption{\small Electron density difference $\Delta$ n(z) between the junctions density and that of its two components(black lines), where the functions attributed to the molecular region are highlighted in green and the running integral over this molecular part is plotted in red for a) s-FDT and b) s-FPT.
 \label{Fig.deltaN}} 
\end{center}
\end{figure*}

In Ref. ~\cite{Verwuste}, where the coverage dependence of FLA for biphenyl-dithiolate monolayers on Au(111) was studied, it was found that the tilt-angle dependence of Pauli repulsion at the interface played a crucial role. In order to investigate whether this is also the case for our molecules we plot in Fig. ~\ref{Fig.deltaN} the electron density difference $\Delta$ n(z) between the density of the junction and of its components in the transport direction and the running integral over its molecular part for the junction geometries of Fig. ~\ref{Fig.conductance_dependence}, where no big differences between these functions for different rotation angles can be identified at the interfaces. This finding further confirms the conclusions from the FLA analysis we presented so far, namely that the level alignment determining the conductance trends in dependence on rotation angle (or junction length) in our article does not rely on interface effects but is defined by the accumulative effects of the respective molecular dipole moments and the partial charging driven by the respective molecular electronegativities, which result in the observed energetic positions of the HOMO in the junction.
\subsection{Summary}\label{paper4:summary}

In this study we investigated the potential use of molecular wires containing ferrocene centers with the same and different anchor groups on each side. We found that the conductance of a proposed set of configurations for molecules s-FDT and s-FPT formed in the junction with two anchoring atoms absorbed to the surface vary in general only within a rather small range in dependence on the rotation angles or type of molecule. The counter intuitive phenomenon that the conductance does not increase while the junction length decreases is explained by Fermi level alignment where we found that the HOMO position for the bended configurations is close to the Fermi level and dominates the conductance.  

We found similar values and patterns of the conductance for both molecules, in the sense that the conductance from junctions where molecules are adsorbed with both anchor sites on the surface have higher conductance values (10$^{-4}$ \texttwelveudash 10$^{-5}$) while a significantly lower conductance value ($\sim$ 10$^{-7}$) is found for the junction geometry formed by a molecule bonded to the tip with one anchor and to the substrate with the other.  
Our study allows experimentalists to relate their measurement data on similar systems to the relative abundance of potential junction geometries. 

\section{Appendix \rom{1}}\label{Appe1}
\subparagraph{Scissor Operator corrections}\label{SO correction}
In order to compare the conductance with experimental results, we need to account for self-interaction errors and image charge effects in DFT calculations with semi-local XC-functionals. We applied a so-called scissor operator (SO) correction according to Quek et al ~\cite{Quek}, which in effect widens the single particle HOMO-LUMO gap and accounts for many body contributions in a very approximative way.
For a molecule adsorbed to the metal surface, SO is defined as $\Sigma_{0}-\Delta\Sigma_{0}$, where $\Sigma_{0}$ is calculated for the highest occupied molecular orbital (HOMO) as
\begin{equation}
 \Sigma_{0}^{o} = -(\varepsilon_{HOMO}+IP),
\end{equation}
where $IP$ and $\varepsilon_{HOMO}$ are the first ionization energy and the HOMO energy with respect to the vacuum potential for the isolated molecule. $IP$ is defined as $IP = E(N-1) - E(N)$ where E(N-1) and E(N) are the respective total energies for the molecule with one positive charge and the neutral case.
For the lowest unoccupied MO (LUMO) $\Sigma_{0}^{u}$ is calculated as
\begin{equation}
 \Sigma_{0}^{u} = -(\varepsilon_{LUMO}+EA),
\end{equation}
where $EA$ is the electron affinity of the isolated molecule, which is calculated as $EA = E(N) - E(N+1)$

The image charge contribution $\Delta\Sigma_{0}$ accounting for screening of the charge by the metal surface we calculate following the recipe by Thygesen et al ~\cite{Thygesen}. This part actually corrects for the polarization of the molecular subsystem and the metal electrodes, and moves the shifted occupied/unoccupied states slightly closer to the Fermi level again, thereby diminishing the original effect of $\Sigma_{0}$. The molecular states obtained by a subdiagonalization of the transport Hamiltonian are then adjusted by $\Sigma_{0}^{o} - \Delta\Sigma_{0}^{o}$ for all occupied states and $\Sigma_{0}^{u} - \Delta\Sigma_{0}^{u}$ for all unoccupied states.
In this way the molecular HOMO-LUMO gap is enlarged by pushing occupied states down and unoccupied states up in energy.

\section{Appendix \rom{2}}\label{Appe2}
\subparagraph{Vacuum Level Alignment}
The alignment of the vacuum level is necessary for defining the same energy reference for molecular eigenenergies and the Fermi level of metallic electrodes in order to relate the MOs to $E_{F}$ when both are obtained from separate DFT calculations for the two different components of the junction. 
This can be done in the following way:
i) In order to obtain the molecular eigenenergies with respect to the vacuum potential, a calculation of the isolated molecule in the same unit cell as the junction structure needs to be performed, where the obtained MO energies are defined as:
\begin{equation}
\varepsilon_{i,vac} = \varepsilon_{i,DFT} - V_{mol,vac}
\end{equation}
with $\epsilon_{i,DFT}$ the molecular eigenvalues from a DFT calculation with an arbitrary code -dependent definition of the energy reference and $V_{mol,vac}$ obtained by averaging the electrostatic potential from the same DFT calculation within the vacuum region.

ii) For the calculation of the metal Fermi energy relative to the vacuum potential $V_{F,vac}$, one needs to perform a DFT calculation of the metal electrodes without the molecule in the same unit cell as the junction structure. The vacuum potential can then
be determined as the electrostatic potential from that calculation in the gap between the two surfaces and the Fermi level can now be determined relative to this vacuum level as:
\begin{equation}
\varepsilon_{F,vac} = \varepsilon_{F,DFT} - V_{F,vac}
\end{equation}
where again the energy reference for $\varepsilon_{F,DFT}$ can be arbitrary and in general code-dependent.

iii) Now we align the MO eigenenergies to the metal's Fermi level by forming the difference 
\begin{equation}
\epsilon_{i,F} = (\epsilon_{i,vac}) -(\varepsilon_{F,vac}).
\end{equation}

\begin{acknowledgments}
Both authors XZ and RS have been supported by the Austrian Science Fund FWF (project number No. P27272). We are indebted to the Vienna Scientific Cluster VSC, whose computing facilities were used to perform all calculations presented in this paper (project No. 70671). We gratefully acknowledge helpful discussions with Tim Albrecht and Mario Lemmer.   
\end{acknowledgments}

\bibliographystyle{apsrev}

\bibliographystyle{apsrev}

\end{document}